\documentclass[traditabstract]{aa}
\usepackage{graphicx}
\usepackage{txfonts}
\usepackage{rotating}
\usepackage{lscape}
\usepackage{longtable}
\usepackage{natbib}
\begin{document}

\title{X-ray, optical and  infrared investigation of the candidate Supergiant Fast X-ray Transient  IGR~J18462$-$0223}

\authorrunning{V.~Sguera}

\author{V. Sguera\inst{1} \and S. P. Drave\inst{2} \and L. Sidoli\inst{3} \and N. Masetti\inst{1} 
\and  R. Landi\inst{1} \and  A. J. Bird\inst{2} \and A. Bazzano\inst{4}}
 
\offprints{sguera@iasfbo.inaf.it}
\institute{INAF, Istituto di Astrofisica Spaziale e Fisica Cosmica, Via Gobetti 101, I-40129 Bologna, Italy \and
School of Physics and Astronomy, University of Southampton, University Road, Southampton, SO17 1BJ, UK  \and
INAF, Istituto di Astrofisica Spaziale e Fisica Cosmica, Via E. Bassini  15, I-20133 Milano, Italy  \and
INAF, Istituto di Astrofisica e Planetologia Spaziali, Via Fosso del Cavaliere 100, I-00133 Rome, Italy}

\date{Received 23 Nov 2012  / accepted 6 May 2013}

      
\abstract
{We report on  a broad-band X-ray study (0.5--60 keV) of the poorly known candidate Supergiant Fast X-ray Transient (SFXT)  
IGR~J18462$-$0223, and on optical  and near-infrared (NIR) followup observations of field objects.
The out-of-outburst X-ray state has been investigated for the first time with archival {\it INTEGRAL}/IBIS, {\it ASCA}, 
{\it Chandra} and {\it Swift}/XRT observations.
This allowed us to place stringent 3$\sigma$ upper limits on the soft (0.5--10 keV) and hard (18--60 keV) X-ray emission 
of 2.9$\times$10$^{-13}$ erg cm$^{-2}$ s$^{-1}$ and 8$\times$10$^{-12}$ erg cm$^{-2}$ s$^{-1}$, respectively; the source 
was also detected during an intermediate soft X-ray state with flux equal to 1.6$\times$10$^{-11}$ erg cm$^{-2}$ s$^{-1}$ 
(0.5--10 keV). In addition, we report on the {\it INTEGRAL}/IBIS discovery of three fast hard X-ray flares (18--60 keV) 
having a duration in the range 1--12 hours: the flaring behavior was also investigated in soft X-rays (3--10 keV)
with archival {\it INTEGRAL}/JEM--X observations. The  duty cycle (1.2\%) and the dynamic ranges ($>$ 1,380 and $>$ 190 
in the energy bands 0.5--10 keV and 18--60 keV, respectively) were measured for the first time.  Archival UKIDSS
$JHK$ NIR data, together with our deep $R$-band imaging of the field, unveiled a single, very red object inside the 
intersection of the {\it Swift}/XRT and {\it XMM-Newton} error circles: this source has optical/NIR photometric properties 
compatible with a very heavily absorbed blue supergiant located at $\sim$ 11 kpc, thus being a strong candidate counterpart for IGR J18462$-$0223. 
NIR spectroscopy is advised to confirm the association. Finally, a hint of a possible orbital period was found at 
$\sim$2.13 days. If confirmed by further studies, this would make IGR~J18462$-$0223 the SFXT with the shortest orbital 
period among the currently known systems.

}

\maketitle

 \section{Introduction}
The {\it INTEGRAL} observatory (Winkler et al. 2003), launched in October 2002,  has opened a new era in the study 
of Supergiant High Mass X-ray Binaries (SGXBs). In fact the Galactic Plane monitoring performed by 
the {\it INTEGRAL}/IBIS instrument (Ubertini et al. 2003) led to the discovery of a new 
sub-class of SGXBs named as Supergiant Fast X-ray Transients (SFXTs).
SFXTs host an OB blue supergiant star  as the companion donor (Negueruela et al. 2006) and display a
peculiar fast X-ray transient behavior lasting  typically a  few hours and no longer than a few days at most (Sguera et al. 2005, 2006). 
The typical  dynamic ranges, from X-ray outbursts ($L_{{\rm X}}$  $\sim$ 10$^{36}$ $-$ 10$^{37}$ erg s$^{-1}$) to lowest level of X-ray emission, 
are of the order of  10$^3$--10$^5$.  Such peculiar fast X-ray transient  behavior is at odds 
with the X-ray characteristics  of their  historical parent population of wind-fed SGXBs
which are detected as  bright  persistent X-ray sources with typical $L_{{\rm X}}$  of $\sim$10$^{36}$erg s$^{-1}$. 
 
Although SFXT hunting is not an easy task, in a few years $\sim$ 10 firm SFXTs have been 
reported in the literature (see list in Grebenev 2010) plus a similar number of candidate SFXTs (e.g. Sidoli et al. 2012, Fiocchi et al. 2010).
 The latter are still unidentified X-ray sources  displaying  a fast X-ray
transient behavior and a X-ray spectral shape strongly resembling those from firm SFXTs. However only infrared/optical 
follow-up observations  of the pinpointed counterpart can provide  a firm classification as SFXT, or its dismissal (e.g. AX J1749.1$-$2733, 
Zurita Heras \& Chaty (2008)), through the 
spectroscopical identification of an early-type supergiant star.
One of the main aims of the current studies on SFXTs is to collect more detailed spectral/temporal 
information on poorly known candidate SFXTs, in order to confirm their nature and so increase the sample 
of established objects. This is mandatory for a population study, e.g. to establish if SFXTs are a homogeneous class or 
display a variety of different X-ray characteristics. It  is also important to unveil the evolutionary paths and formation rate of SFXTs in our Galaxy as well as their accretion mechanisms, which are still largely unknown. 

Among the candidate SFXTs, IGR~J18462$-$0223 is one of the poorest studied
to date. In the hard X-ray band (20--60 keV), the only information reported in the literature   
come from  two fast X-ray outbursts detected by  {\it INTEGRAL}/IBIS (Grebenev \& Sunyaev 2010),  
one in April 2006 (duration and average  flux of 1 hour and 65 mCrab, respectively) and  another one in October 2007 (5 hours and 35 mCrab). 
During  both outbursts  the average 20--60 keV X-ray spectrum is well fit by a power law ($\Gamma$ $\sim$ 2.5) 
or alternatively a bremsstrahlung  model (kT $\sim$ 40 keV).  In addition, a possible cyclotron resonance feature
around 25 keV was reported by Grebenev \& Sunyaev (2010), although the authors could not  draw 
any conclusion on the genuine existence of such feature. As for the soft  X-ray band (0.5--10 keV),  
IGR~J18462$-$0223 was observed by {\it XMM-Newton} for 32 ks on April 2011 (Bodaghee et al. 2012).  Its  X-ray spectrum 
is well modeled by an absorbed power law which  yields a large absorbing column density ($N_{{\rm H}}$$\sim$ 3$\times$10$^{23}$ cm$^{-2}$) 
and a  photon index equal to $\Gamma$ $\sim$ 1.5.  The unabsorbed X-ray flux  is  3.6$\times$10$^{-11}$ erg cm$^{-2}$ s$^{-1}$ (0.5--10 keV). 
An iron line at 6.4 keV and a break in the power law at 3 keV are also present in the  X-ray spectrum. These  spectral characteristics 
are typical of wind-fed accreting X-ray pulsars and in fact  X-ray pulsations were also discovered  at $\sim$ 997 seconds (Bodaghee et al. 2012). 
The catalogued 2MASS infrared  source nearest to the {\it XMM-Newton}  error circle (2$\hbox{$.\!\!^{\prime\prime}$}$5  radius) is 2MASS J18461279--0222261/USNO B--1.0 0876--0579765 which is about 3$\hbox{$.\!\!^{\prime\prime}$}$4 away, i.e. outside of it (Bodaghee et al. 2012).

Here we report on both soft  (0.5--10 keV) and hard (18--60 keV) X-ray properties   of IGR~J18462$-$0223
during the out-of-outburst state,  as obtained from archival {\it ASCA}, {\it Chandra}, {\it Swift}/XRT and 
{\it INTEGRAL}/IBIS observations.  
We also report on {\it INTEGRAL}/IBIS and {\it INTEGRAL}/JEM--X spectral and timing analysis of three newly 
discovered  fast X-ray outbursts.
 Next, we investigated  the IBIS/ISGRI and {\it RXTE}/ASM long-term monitoring light curves of the source 
searching for orbital periodicities.
Moreover, to explore the field of IGR~J18462$-$0223 and investigate its  longer-wavelength counterpart, we 
acquired optical spectra of 2MASS J18461279$-$0222261 and deep $R$-band imaging of the source field with the 3.58m 
Telescopio Nazionale  Galileo located in Canary Islands (Spain) and with the 1.5m Cassini telescope located in 
Loiano (Italy), respectively;  we also retrieved near-infrared (NIR) images of the field collected within the UKIDSS 
Galactic Plane Survey (Lucas et al. 2008).

\begin{table*}
\begin{center}
\caption {Summary of the characteristics of all hard X-ray outbursts detected by {\it INTEGRAL}/IBIS from IGR~J18462$-$0223  to date.
 The table lists the date of their peak emission, energy band of the detection, 
approximate duration  and significance detection of the entire flaring activity,  
X-ray flux  at the peak,  average flux and photon index of the power law spectrum. Outbursts number 2 , 4, 5 are newly discovered while outbursts number 1 and  3,   
indicated by the symbol (*), have been already reported in the literature by 
Grebenev \& Sunyaev (2010).  $\ddagger$ = outburst  also detected by {\it INTEGRAL}/JEM--X.} 
\label{tab:main_outbursts} 
\begin{tabular}{lccccccc}
\hline
\hline   
N. & peak-date  & energy band                &  duration       & significance         &  peak-flux      & average flux                         &    $\Gamma$                    \\
    &  (MJD)         &     (keV)           &   (hours)                       & ( $\sigma$)   &        (erg cm$^{-2}$ s$^{-1}$)   &  (erg cm$^{-2}$ s$^{-1}$)   &      (power law)  \\
  \hline    
1 (*) $\ddagger$  &  $\sim$ 53853.40    & 18--60            & $\sim$ 1             & 7.8  &    (1.5$\pm$0.2$)\times$10$^{-9}$ &   4.0$\times$10$^{-10}$          &   2.2$^{+0.7}_{-0.7}$      \\                                                                          
2      &$\sim$ 53981.20     & 18--60               &$\sim$ 12             & 8.2                   & (2.6$\pm$0.7)$\times$10$^{-10}$   &9.7$\times$10$^{-11}$ &  2.0$^{+0.7}_{-0.7}$   \\    
3 (*) &  $\sim$ 54385.75      & 18--60             & $\sim$ 5            &  11.1             &  (6.4$\pm$1.4$)\times$10$^{-10}$     &   2.6$\times$10$^{-10}$      &     2.4$^{+0.6}_{-0.6}$     \\
4      & $\sim$ 54519.89     &18--60               & $\sim$ 1              & 5.7                  & (7.0$\pm$1.7$)\times$10$^{-10}$   &   2.7$\times$10$^{-10}$    &          \\   
5      & $\sim$   54909.08    & 18--60             & $\sim$ 4           & 5.7                     &   (7.0$\pm$2.1$)\times$10$^{-10}$ &   1.1$\times$10$^{-10}$       &    \\  
\hline
\end{tabular}
\end{center}
\end{table*}

\begin{table*}
\begin{center}
\caption {Summary of characteristics of all soft X-ray  observations targeted at IGR J18462$-$0223 to date.
The table lists the date of the observation, energy band and X-ray satellite performing the observation, exposure time, unabsorbed average flux,  X-ray state of the source, photon index and total absorption of the power law spectrum.  $\ddagger$ = 3$\sigma$ upper limit.  The observation indicated by the symbol (*) has been already reported in the literature by 
Bodaghee et al. 2012.}

\label{tab:main_outbursts} 
\begin{tabular}{lccccccc}
\hline
\hline   
 date   & energy band            &X-ray satellite        &   exposure                          &   average flux            & X-ray state                                               &    $\Gamma$   &    $N_{{\rm H}}$            \\
      (MJD)         &     (keV)            &                       &    (ks)              & (erg cm$^{-2}$ s$^{-1}$)      &                        &      (power law)   & (10$^{22}$, cm$^{-2}$) \\
  \hline           
 50561.00            &  0.5--10   & ASCA                    &   22            &  $<$ 2.9$\times$10$^{-13}$   $\ddagger$    & quiescence           &                     &                            \\                                                                    
 53853.35        & 3--10     &  INTEGRAL/JEM--X  &     2          & 1.0$\times$10$^{-10}$                & outburst          & 0.98$\pm$0.55                      &      \\  
 54766.65        &  0.5--10   & Chandra                  &   1              &  $<$ 7.1$\times$10$^{-12}$ $\ddagger$  & quiescence    &                               &                  \\
 55669.37 (*)       & 0.5--10     &  XMM                    &    32              & 3.6$\times$10$^{-11}$                & intermediate          & 1.5$\pm$0.1           & 28$\pm$1                \\    
 55877.80             &  0.5--10   & Swift/XRT             &    2          & 2.7$\times$10$^{-11}$               & intermediate               &  1.7$\pm$0.2    &  19.7$^{+5.7}_{-4.6}$ \\
\hline
\end{tabular}
\end{center}
\end{table*}

\section{Data Analysis}
For the {\it INTEGRAL} study, we used all the public data collected with IBIS (Ubertini et al. 2003)
from February 2003 to May 2011. In particular, the data set consists of 4742 pointings or Science Windows 
(ScWs, $\sim$ 2000 seconds duration) where IGR J18462$-$0223 was within 12$^\circ$  from the centre of the 
instruments field of view (FoV), for a total on-source time of 6.96 Ms.  A 12$^\circ$ limit was applied because the off-axis response of 
IBIS (whose FoV is 30$^\circ$$\times$30$^\circ$) is not well modelled  at large angles and consequently  may introduce a systematic 
error in the measurement of source fluxes. The data reduction was carried out with the 
release 9.0 of the Offline Scientific Analysis software (OSA). IBIS/ISGRI (Lebrun et al. 2003) images for each pointing were generated in the energy band 18--60 keV.

We also used unpublished  soft X-ray observations  of  IGR J18462$-$0223 (0.5--10 keV)  from the {\it ASCA}, {\it Chandra} and {\it Swift}/XRT archive.

Through the paper, all spectral analysis was performed  using \emph{Xspec}  version 11.3; 
uncertainties are given at the 90\% confidence level for one single parameter of interest.

\begin{figure}
\begin{center}
\includegraphics[height=8cm,width=5.5cm, angle=270]{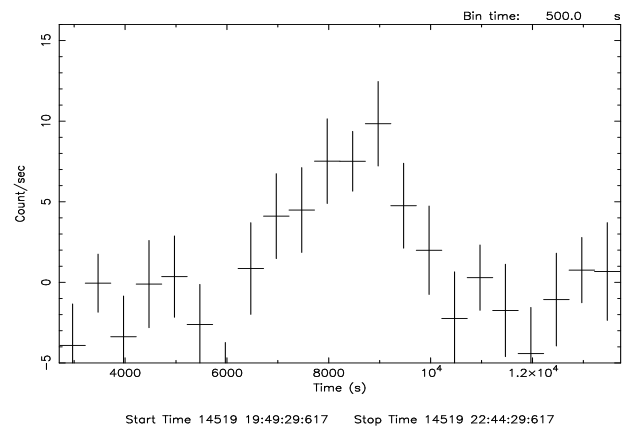}
\caption{IBIS/ISGRI light curve (18--60 keV) of a newly discovered fast X-ray outburst from IGR J18462$-$0223 (N. 4 in Table 1). The bin time is 500 seconds.} 
\end{center}
\end{figure} 

\section{INTEGRAL results}
\subsection{IBIS/ISGRI}
\subsubsection{Outbursts}
An analysis at the ScW level of all the deconvolved IBIS/ISGRI shadowgrams (18--60 keV) 
has been performed to search for new fast X-ray flares  from IGR J18462$-$0223 detected with a significance greater than
at least 5$\sigma$ and lasting from few hours to a few days.  As a result,  we report three newly discovered fast X-ray outbursts which are listed in Table 1 together  with the date of the peak emission, approximate duration,  outburst detection significance, flux at outburst peak, average flux and  photon index of the power law spectrum. Since it was not possible to perform a meaningful spectral analysis for the outbursts having the lowest significance detection (N. 4 and 5 in Table 1),  we assumed a Crab-like energy spectrum (photon index equal to 2.1) to calculate their fluxes.

The strongest as well as shortest newly discovered outburst  (N. 4) was detected at 5.7$\sigma$ level during only one
ScW lasting $\sim$ 1 hour,  at an  off-axis angle of  about 6.6 degrees. Fig. 1  shows the IBIS/ISGRI light curve of the entire outburst activity with a bin time of 
500 seconds. It reached a peak flux of 54$\pm$13 mCrab  or (7.0$\pm$1.7)$\times$10$^{-10}$ erg cm$^{-2}$ s$^{-1}$  (18--60 keV) on 4 March 2008,
its average flux during the entire outburst duration is equal to  2.7$\times$10$^{-10}$ erg cm$^{-2}$ s$^{-1}$. 
The limited statistics (5.7$\sigma$ detection) prevent us from a meaningful spectral analysis. 
Another newly discovered fast X-ray outburst (N. 2) reached a 18--60 keV 
peak flux of 20$\pm$6 mCrab or (2.6$\pm$0.7)$\times$10$^{-10}$ erg cm$^{-2}$ s$^{-1}$ on 
3 September 2003.  The total outburst duration was $\sim$ 12 hours for a detection of 8.2$\sigma$. 
Its 18--60 keV spectrum is well fitted by a simple power law model with $\Gamma$=2.0$^{+0.7}_{-0.7}$  ($\chi^{2}_{\nu}$=0.96, 15 d.o.f.),
the average flux (18--60 keV) is equal to  9.7$\times$10$^{-11}$ erg cm$^{-2}$ s$^{-1}$.
Finally, the last newly discovered fast X-ray outburst (N. 5) was detected on 19 March 2009 at 5.7$\sigma$ level with a duration of 
$\sim$ 4 hours. It  reached a 18--60 keV peak flux of  54$\pm$16 mCrab or (7.0$\pm$2.1)$\times$10$^{-10}$ erg cm$^{-2}$ s$^{-1}$ while the average flux during the entire  outburst  is equal to  1.1$\times$10$^{-10}$ erg cm$^{-2}$ s$^{-1}$.  As in the case of outburst N. 4,  the limited statistics prevent us from a meaningful spectral analysis.

Table 1 provides a summary of the characteristics
of all hard X-ray outbursts detected by {\it INTEGRAL}/IBIS from IGR~J18462$-$0223  to date. 
We note that the  newly discovered X-ray outbursts (N. 2, 4, 5) are similar,  in term of duration and spectral shape,  to those previously reported  in the literature  (N. 1, 3)  by  Grebenev \& Sunyaev (2010). 

Grebenev \& Sunyaev (2010) suggested the presence of a possible cyclotron resonance feature
at $\sim$ 26 keV  in the summed X-ray spectra (20--60 keV) of two already published outbursts (N. 1 and 3 in Table 1). 
Bearing in mind that the  X-ray  outbursts  N. 1, 2 and 3 in Table 1 have a very similar spectral shape within the large uncertainties, 
we summed them up  with the aim of improving the statistics and  investigate the presence of a possible cyclotron absorbing feature.
 A power law ($\Gamma$=2.1$\pm$0.5, $\chi^{2}_{\nu}$=1.3, 15 d.o.f.) provided a reasonable  description of the average spectrum 18--60 keV. 
We therefore added to the power law model a cyclotron  absorption line (cyclabs in XSPEC notation), 
however  no such feature was statistically required by the data. 
Further studies with much better statistics are needed to  fully confirm or reject the genuine existence of such feature.

\subsubsection{Out-of-outburst emission}
The out-of-outburst hard X-ray  behaviour of  IGR~J18462$-$0223 (18--60  keV)  is totally unknown.
We searched the entire IBIS/ISGRI public data archive for
pointings where IGR J18462$-$0223 was within the fully coded FoV 
of IBIS (9$^\circ$$\times$9$^\circ$). Subsequently  we excluded
those individual ScWs during which the source was in outburst. We collected a total
of 371 ScWs which were used to generate a mosaic significance
map in the 18--60 keV band for a total exposure of 712 ks.  
IGR~J18462$-$0223 was not detected and we inferred a 18--60 keV 3$\sigma$ upper limit of 0.6 
mCrab or 8$\times$10$^{-12}$ erg cm$^{-2}$ s$^{-1}$  (again by assuming a Crab-like energy spectrum with photon index equal to  2.1).
If we assume the highest source flux to take place
in an outburst during its peak (1.5$\times$10$^{-9}$ erg cm$^{-2}$ s$^{-1}$) ,  as measured by IBIS/ISGRI in the energy range 18--60 keV,  
we can derive a dynamic range of $>$ 190.

\subsection{JEM--X}
The X-Ray Monitor JEM--X  (Lund et al. 2003)  on board the
{\it INTEGRAL} satellite performs  observations simultaneously with IBIS/ISGRI, providing images in the energy band 3--35 keV
with a 13$^\circ$.2 diameter partially coded FoV (PCFoV).  Images from JEM--X (3--20 keV) were created for all outbursts reported in Table 1. 
Because of  the  much smaller JEM--X PCFoV compared to the IBIS one (30$^\circ$$\times$30$^\circ$), only in one case  was the source inside the JEM--X PCFoV 
so that it was possible to extract a spectrum  and a X-ray light curve. In fact, during the outburst that occurred in April 2006 (N. 1 in Table 1) with $\sim$ 1 hour 
duration, the source was also detected by JEM--X at 5.2$\sigma$ level (3--20 keV) with an effective exposure time of 
$\sim$ 1.7 ks.  From the JEM--X light curve of the X-ray outburst binned with 100 seconds, we estimated a 3--10 keV (3--20 keV) source  peak flux 
equal to 4.0$\pm$0.7  (6.1$\pm$1.1) $\times$10$^{-10}$ erg cm$^{-2}$ s$^{-1}$. We extracted a JEM--X spectrum (3--20 keV) which was reasonably fitted with a simple  power law model ($\chi^{2}_{\nu}$=0.96, 131 d.o.f.) having a hard photon index ($\Gamma$=0.98$\pm$0.55).  Unfortunately the JEM--X data extend down only to 3 keV, i.e. not low enough 
in energy to allow an investigation of the X-ray absorption. The average flux during the X-ray outburst was $\sim$  1.0$\times$10$^{-10}$ erg cm$^{-2}$ s$^{-1}$ (3--10 keV) and 
2.2$\times$10$^{-10}$ erg cm$^{-2}$ s$^{-1}$ (3--20 keV). We also performed the broadband spectral analysis of the simultaneous JEM--X/ISGRI outburst spectrum: a good fit 
was achieved by a power law ($\chi^{2}_{\nu}$=1.09, 139 d.o.f.)  having  $\Gamma$=1.3$\pm$0.5 and a cross-calibration constant of 1.0$^{+1.7}_{-0.6}$. The 3--60 keV  average  flux is  equal to 
5.7$\times$10$^{-10}$ erg cm$^{-2}$ s$^{-1}$.


\section{Archival soft X-ray observations}
\subsection{ASCA}
From archival {\it ASCA} data, we found that on April 1997 IGR~J18462$-$0223   was inside the GIS2 FoV for $\sim$ 22 ks during an observation of the {\it ASCA} Galactic Plane survey  (Sugizaki et al. 2001).  We reduced the data and found that the source  was not detected.
A background count rate of 9.6$^{-4}$ cts s$^{-1}$ was derived and then used it with WEBPIMMS in order to estimate a 0.5--10 keV 
3$\sigma$ upper limit of 2.9$\times$10$^{-13}$ erg cm$^{-2}$ s$^{-1}$ (we assumed the same spectral model of the 
{\it Swift}/XRT observation). If we consider the highest source flux to take place in an outburst as detected by {\it INTEGRAL}/JEM--X  in a similar energy range, then we can infer a dynamic range of $>$ 1,380.

\begin{figure}
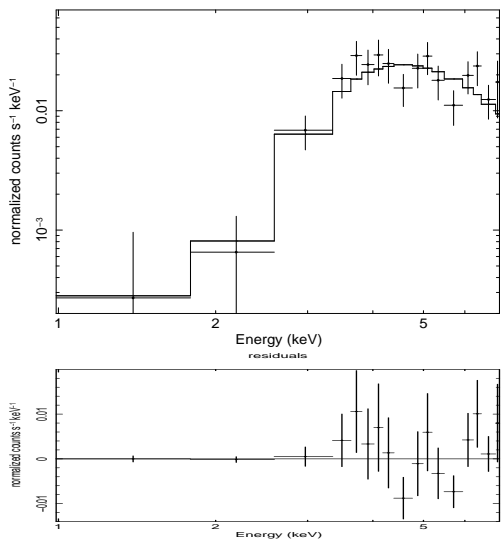

\begin{center}
\includegraphics[height=6.5cm,width=4.5cm, angle=270]{fig2_a.ps}
\includegraphics[height=6.5cm,width=2.5cm, angle=270]{fig2_b.ps}
\caption{{\it Swift}/XRT spectrum of IGR J18462$-$0223 best fitted by an absorbed power law model (top) and relative residuals (bottom).} 
\end{center}
\end{figure}

\begin{figure}
\begin{center}
\includegraphics[height=5.5cm,width=7.5cm]{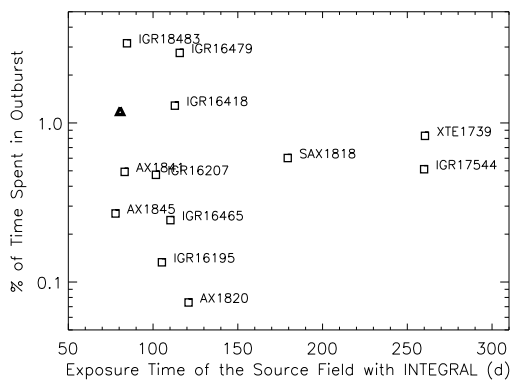}
\caption{Percentage of time spent in outburst vs exposure time (days) for a sample of 13 firm/candidate SFXTs as taken from Ducci et al. (2010). IGR J18462$-$0223 is indicated by means of a black triangle.} 
\end{center}
\end{figure}

 \subsection{{\it Swift}/XRT}
A search in the \emph{Swift}  (Gehrels et al. 2004)  X-ray telescope (XRT, 0.2--10 keV, Burrows et al. 2005) 
data archive revealed that XRT  pointed at IGR J18462$-$0223 (source on-axis)  on 12 November 2011  for a total exposure of $\sim$ 2  ks. 

XRT data reduction was performed using the XRTDAS standard data pipeline 
package ({\sc xrtpipeline} v. 0.12.6), in order to produce screened event files. All data 
were extracted only in the Photon Counting (PC) mode (Hill et al. 2004), adopting the 
standard grade filtering (0--12 for PC) according to the XRT nomenclature. Events for 
spectral analysis were extracted within a circular region of radius 20$^{\prime \prime}$, 
centered on the source position, which encloses about 90\% of the PSF at 1.5 keV (see 
Moretti et al. 2004). The background was taken from various source-free regions close to 
the X-ray source of interest, using circular regions with different radii in order to 
ensure an evenly sampled background. In all cases, the spectra were extracted from the 
corresponding event files using the {\sc XSELECT} software. 
We used v. 013 of the response matrices and created individual ancillary response files \textit{arf} 
using {\sc xrtmkarf v. 0.6.0.} IGR J18462$-$0223 was detected at $\sim$ 13$\sigma$ level, the best determined XRT position is at  ${\rm RA}=18^{\rm h}$46$^{\rm m}$12.84$^{\rm s}$, ${\rm Dec}=-02^{\circ}$22' 29".7 (J2000) with the error radius being 
equal to 2$\hbox{$.\!\!^{\prime\prime}$}$4\footnote{90\% c.l., we used the XRT--UVOT alignment and matching UVOT field sources to the USNO-B1 
catalog, see Evans et al. 2009 and http:/www.swift.ac.uk/user$\textunderscore$objects}. Such coordinates are 2$\hbox{$.\!\!^{\prime\prime}$}$4 from (and compatible with)  the {\it XMM-Newton} position reported by Bodaghee et al. (2012)  with 
an error radius of 2$\hbox{$.\!\!^{\prime\prime}$}$5 (see Fig. 7 and relative discussion in section 6). 

From the  {\it Swift}/XRT data, it was possible to extract a meaningful spectrum only in the energy range 1--7 keV. Below 1 keV  and above 7 keV, the statistics are not good enough to perform a reliable spectral analysis.  Because of the small number of counts,  we used the  Cash  statistic on the unbinned data (Cash 1979). Firstly the spectrum was fitted by an absorbed power law where the absorption was fixed to the Galactic value of 1.85$\times$10$^{22}$ cm$^{-2}$ (Kalberla et al. 2005), however the fit (C-statistics/d.o.f.=496/598) was characterized by a  hard photon index and   the residuals strongly suggested the presence of extra absorption at lower energies. The addition of an intrinsic absorption significantly improved the fit  (C-statistics/d.o.f.=380/597; $\Delta$C=116\footnote{on the basis of the Wilks theorem (1938,1963), Cash demonstrates that $\Delta$C is distributed as $\Delta$$\chi$$^2$, and consequently the confidence levels are determined in the same way of the $\chi$$^2$ statistics}). The best fit parameters are  $\Gamma$=1.7$^{+0.2}_{-0.2}$ and intrinsic $N_{{\rm H}}$=(19.7$^{+5.7}_{-4.6}$) $\times$ 10$^{22}$ cm$^{-2}$ (see spectrum in Fig. 2), values  that are fully compatible within the uncertainties with those obtained with {\it XMM-Newton}  (Bodaghee et al. 2012). In particular,  the 
intrinsic $N_{{\rm H}}$ measured  by both  {\it XMM-Newton}  and {\it Swift}/XRT (i.e. $\sim$ 3$\times$ 10$^{23}$ cm$^{-2}$) is higher than that characterizing other classical SFXTs (i.e.  $\sim$ 10$^{22}$ cm$^{-2}$). 
The {\it Swift}/XRT 0.5--10 keV absorbed (unabsorbed) average flux is 1.6$\times$10$^{-11}$ erg cm$^{-2}$ s$^{-1}$ 
(2.7$\times$10$^{-11}$ erg cm$^{-2}$ s$^{-1}$). We note that this is a flux state similar to that  measured by {\it XMM-Newton}  
(unabsorbed 3.6$\times$10$^{-11}$ erg cm$^{-2}$ s$^{-1}$) in the same energy band.  A light curve  was extracted from the {\it Swift}/XRT data 
and we noted that  the source flux remained constant, within errors, with no sign of flaring activity throughout the duration of the observation.

 \subsection{Chandra}
The ACIS-I detector  onboard {\it Chandra} observed IGR J18462$-$0223 on 27 October 2008 for a total exposure time of $\sim$ 1 ks .
We reduced the data using the latest Ciao software (v 4.4) and calibration file.  No source was detected inside the ACIS-I  FoV and we derived a background count rate of 0.05 cts s$^{-1}$ and then used it with WEBPIMMS in order to estimate a 0.5--10 keV 
3$\sigma$ upper limit of 7.1$\times$10$^{-12}$ erg cm$^{-2}$ s$^{-1}$ (we assumed the same spectral model of the {\it Swift}/XRT observation).

\section{Recurrence time of the outbursts and duty cycle}
\subsection{INTEGRAL}
The sky region of IGR J18462$-$0223 has been covered by  {\it INTEGRAL}/IBIS  observations
for a total exposure time of $\sim$ 7 Ms.  {\it INTEGRAL}/IBIS monitoring could be considered as a
sporadic sampling of the light curve with a resolution of 2000 seconds  over a baseline of $\sim$ 8 years. 
We calculated the duty cycle of IGR J18462$-$0223 (i.e. the fraction of time spent in bright outbursts with respect to the total observational time)
following the same criteria adopted by Ducci et al. (2010),  i.e. we considered the duration of only the 
bright outbursts detected with a significance level greater than at least 5$\sigma$.  As result  the duty cycle is
 $\sim$ 1.2\%,  in Fig. 3  we compare it with those of a sample of 
firm/candidate SFXTs analyzed by Ducci et al.  (2010). We note that it is  in the range  0.1\%--3\% found by Ducci et al. (2010).

It is  intriguing to note  that all the  outbursts detected by {\it INTEGRAL} to date  (see Table 1)  are spaced by multiples of $\sim$ 2  days.  In fact if we assume  the occurrence time  of the nth outburst as given by T$_n$=T$_0$+(nT$_{orb}$), where T$_0$=53853.4 MJD is the time  of the first ever detected outburst,  then we found that T$_{orb}$=2.13  days is the longest value able to account for the peak occurrence  of  all the detected outbursts.   In fact,  the  times predicted by the T$_{orb}$=2.13  days periodicity (53981.20 at cycle n=60, 54385.90 at n=250, 54520.09 at n=313,  54909.88 at n=496)
are in very good agreement with those observed by {\it INTEGRAL} at the peak  (see Table 1)  if we consider the outbursts duration 
of the order of hours. Such hint of a possible 2.13 days periodicity  could be interpreted as the likely  orbital period of the binary system.
As next step, in order to search for any evidence of periodicity by using a proper analysis, 
we investigated the IBIS/ISGRI  long-term light curve  with the Lomb--Scargle periodogram method by means of the fast implementation
of Press \& Rybicki (1989) and Scargle (1982).  No signal having a significance level greater than at least 90\% is seen in the periodogram 
at any period, in particular at  $\sim$  2.13 days.

\subsection{RXTE/ASM}
The all sky  instrument ASM (Levine et al. 1996), onboard the {\it RXTE} satellite,  was an X-ray monitor consisting of three Scanning Shadow Cameras (SSCs)  each with a position  sensitive proportional counter.  
ASM measured the counts from  IGR J18462$-$0223 with one of the three SSCs up to several times per  day during $\sim$ 90 seconds long dwells. We downloaded from the ASM team web page (http://xte.mit.edu/ASM$\textunderscore$lc.html)  the   long-term monitoring light curve (2--10 keV)  covering the period from  Jan 1996 to Dec 2009.

\begin{figure}
\begin{center}
\includegraphics[height=5cm,width=7cm]{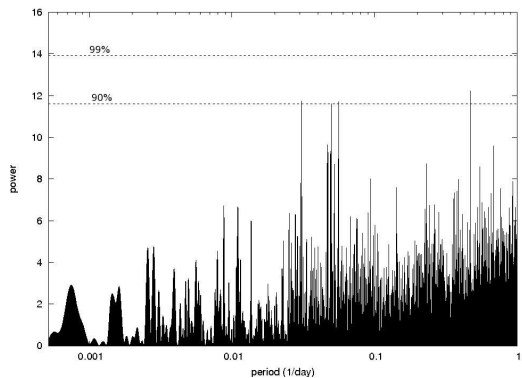}
\caption{Lomb--Scargle periodogram of  IGR J18462$-$0223 {\it RXTE}/ASM dwell by dwell light curve.  The two dotted lines represent the significance level of 90\%  and 99\%, respectively.  No significant signal is detected above 99\%,  the strongest peak is slightly above the 90\% level and it corresponds to a period of 2.1378 days.} 
\end{center}
\end{figure}    

 \begin{figure}
\begin{center}
\includegraphics[height=5cm,width=7cm]{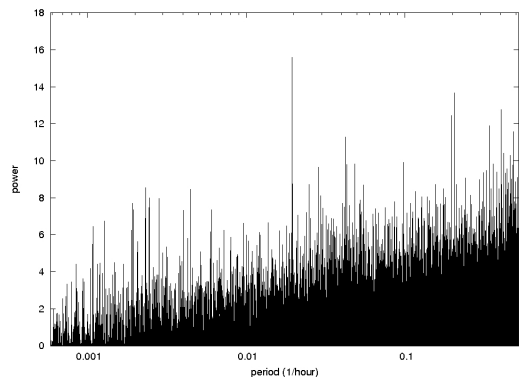}
\caption{Fast Fourier transform periodogram of  IGR J18462$-$0223 {\it RXTE}/ASM dwell by dwell light curve as  downloaded  from the ASM team web page 
(http://xte.mit.edu/ASM$\textunderscore$lc.html) and obtained by using the standard criteria mentioned there.  The strongest peak corresponds to a period of 2.1377 days.} 
\end{center}
\end{figure}    

We  investigated with Lomb--Scargle the  long-term monitoring  {\it RXTE}/ASM dwell by dwell light curve.  
The strongest peak is detected slightly above a significance level of  90\% (see Fig. 4), it corresponds to a period of $\sim$  2.1378 days  which is
in  agreement with the previously inferred value of 2.13 days. Other peaks are also present in the periodogram although they do not exceed  the 90\% significance level.   While this could appear unusual, we note that it is not the first time that a period has been detected
in the light curve of one X-ray instrument and not in that of others (e.g. Drave et al. 2011, Clark et al. 2010). In fact, we note 
that the {\it RXTE}/ASM long-term monitoring light curve has a much longer baseline (14 years)
than IBIS/ISGRI (8 years), in addition ASM measure the source count rate up to several times per day so the
ASM light curve is well sampled throughout the 14 years, which is not the case of the IBIS/ISGRI light curve. This
could possibly explain why an hint of periodicity has been observed only in the {\it RXTE}/ASM periodogram.

As crosscheck, we also downloaded  from the ASM team web page 
the fast Fourier transform power density spectrum of IGR J18462$-$0223 as obtained from the ASM team by using the standard criteria mentioned there 
(i.e. by averaging the light curves into 0.05 day bins, subtracting the mean value of all inhabited bins from each inhabited bin, and taking a fast Fourier transform, 
the resulting power in each frequency bin is divided by the average power over the whole frequency range). 
As it can be seen from Fig 5, again the strongest peak in the periodogram corresponds to a period of $\sim$  2.1377 days. Unfortunately we cannot claim a secure detection and  draw any 
firm conclusion on the genuine presence of a  2.13 days  periodicity in the {\it RXTE}/ASM data  because of the low statistical 
significance ($\sim$  90\%) of  the tentative detected  periodicity.

\section{Search for the optical/near-infrared counterpart}

To explore the field of IGR J18462$-$0223 and investigate its proposed 
counterpart, we followed several approaches.

First, in order to conclusively exclude as the optical counterpart the 
nearest catalogued IR source 2MASS J18461279$-$0222261 mentioned by Bodaghee 
et al. (2012), starting at 02:49 UT on 2012 May 31 we acquired two 20-min 
optical spectra of it with the 3.58m Telescopio Nazionale Galileo, located 
in the Canary Islands (Spain) and equipped with the DoLoRes istrument plus 
LR-B grism and a slit of width 1$\farcs$5.
These spectra were reduced following a standard procedure (Horne 1986) 
within IRAF\footnote{IRAF is the Image Analysis and Reduction Facility made 
available to the astronomical community by the National Optical Astronomy 
Observatories, which are operated by AURA, Inc., under contract with the 
U.S. National Science Foundation. It is available at http://iraf.noao.edu/}.
As expected, given the position of 2MASS J18461279$-$0222261 with respect to 
the {\it XMM-Newton} and {\it Swift} positions of IGR J18462$-$0223 (see 
also Fig. 6), its optical spectrum does not show any peculiar feature and 
appears typical of a very reddened  G-type star.

Next, we performed deep optical imaging on the {\it XMM-Newton}
and {\it Swift}/XRT error circles to look for the presence of fainter
objects within them. To this purpose we acquired two 20-min $R$-band
exposures of the IGR J18462$-$0223 field with the 1.5m `Cassini' telescope
located in Loiano (Italy). The telescope carries the BFOSC instrument,
with an EEV CCD which allows the coverage of a 13$'$$\times$12$\farcm$6 
field with a scale of 0$\farcs$58/pixel. Observations were performed 
starting at 21:29 UT of 2012 July 27, under an average seeing of 1$\farcs$9.

The scientific frames were corrected for bias and flat field and then 
stacked together to increase their S/N ratio, again following a standard 
procedure. To get an estimate of the image depth, we performed a photometric 
study of it. Owing to the field crowdedness (see Fig. 6), we chose standard 
point spread function (PSF) fitting technique by using the PSF-fitting 
algorithm of the DAOPHOT II image data-analysis package (Stetson 1987) 
running within MIDAS\footnote{MIDAS (Munich Image Data Analysis System) 
is developed, distributed and maintained by the European Southern 
Observatory and is available at http://www.eso.org/sci/software/esomidas/}.

The image thus acquired was then also processed  with the 
GAIA/Starlink\footnote{available at
http://star-www.dur.ac.uk/\~{}pdraper/gaia/gaia.html} package to obtain an 
astrometric solution based on 30 USNO-A2.0\footnote{The USNO-A2.0 catalogue is 
available at http://archive.eso.org/skycat/servers/usnoa} reference stars 
in the field of IGR J18462$-$0223. The conservative error in the optical 
position is 0$\farcs$58, which was added in quadrature to the systematic 
error in the USNO catalogue (0$\farcs$25 according to Assafin et al. 2001 
and Deutsch 1999).

The final 1$\sigma$ uncertainty in the astrometric solution of the image 
is thus 0$\farcs$63. Once we had determined the astrometry of the image, 
we superimposed it with the {\it XMM-Newton} and {\it Swift} X-ray error 
circles. The result is reported in Fig. 6, upper panel: no objects are present 
within the {\it XMM-Newton} error circle down to a 3-$\sigma$ magnitude 
limit\footnote{ We stress that the $R$-band magnitude calibration was 
obtained using USNO-A2.0 stars in the field, which are known to have systematic 
uncertainties of up to a few tenths of magnitude in some cases (see Masetti 
et al. 2003), thus the magnitudes reported above may suffer from a 
systematic shift of this amount.}
of $R <$ 21.5, while an object lying southeast of 2MASS J18461279-0222261 and 
of magnitude $R$ = 18.76$\pm$0.04 is contained in the {\it Swift} 
positional uncertainty. This source has coordinates (J2000) RA = 18$^{\rm h}$ 
46$^{\rm m}$ 12$\fs$90, Dec = $-$02$^{\circ}$ 22$'$ 28$\farcs$8.

Finally, to better explore this source and the NIR content of the two X--ray 
error circles we searched the UKIDSS Galactic Plane Survey (Lucas et al. 2008) 
for NIR images of the field (see Fig. 6, lower panel). We found $JHK$ frames 
acquired on 12 June 2006 and we clearly detect the above object at magnitudes 
$J$ = 15.397$\pm$0.005, $H$ = 14.760$\pm$0.004 and $K$ = 14.341$\pm$0.009. 
These values and the associated colors are actually not typical of a reddened 
blue giant counterpart of a SFXT, which is expected to be much brighter in 
the NIR. We thus consider this source also as an unlikely counterpart for 
IGR J18462$-$0223.

In the UKIDSS images 
we moreover detected a number of other objects within or close the {\it Swift} 
and the {\it XMM-Newton} X--ray error circles (as can be seen in the lower panel 
of Fig. 6); however, a more careful look shows just one single NIR object within 
their intersection: this source appears rather bright in the $K$-band image and 
is undetected in the $J$-band one. 
Its UKIDSS coordinates (J2000) are RA = 18$^{\rm h}$ 46$^{\rm m}$ 12$\fs$757, 
Dec = $-$02$^{\circ}$ 22$'$ 28$\farcs$48 (the uncertainty is about 0$\farcs$1; 
Lucas et al. 2008), and the NIR magnitudes are $J >$ 19.5, $H$ = 15.714$\pm$0.010 
and $K$ = 13.178$\pm$0.003 (for the $J$-band we used the survey limit as reported 
in Casali et al. 2007). 
We remark that the UKIDSS astrometry is fully consistent with the X--ray and the 
optical ones within uncertainties, and that the photometry is as well consistent 
with the 2MASS one (Skrutskie et al. 2006) of isolated field stars.
We also notice that this source is positionally coincident (within 0$\farcs$25)
with the mid-IR GLIMPSE catalogue source G030.2231+00.0791 (Benjamin et al. 2003).

All of the above strongly points out that this very  source, lying within the
intersection of the two X--ray positional uncertainties, might be a very reddened blue supergiant
and the actual counterpart of IGR J18462$-$0223. As a corollary, this finding indicates
that the two X--ray positions are not in disagreement but rather allowed, by being
considered together, the discovery of the NIR counterpart of this hard X--ray object.

\begin{figure}
\begin{center}
\includegraphics[height=8.5cm,width=8.5cm,angle=0]{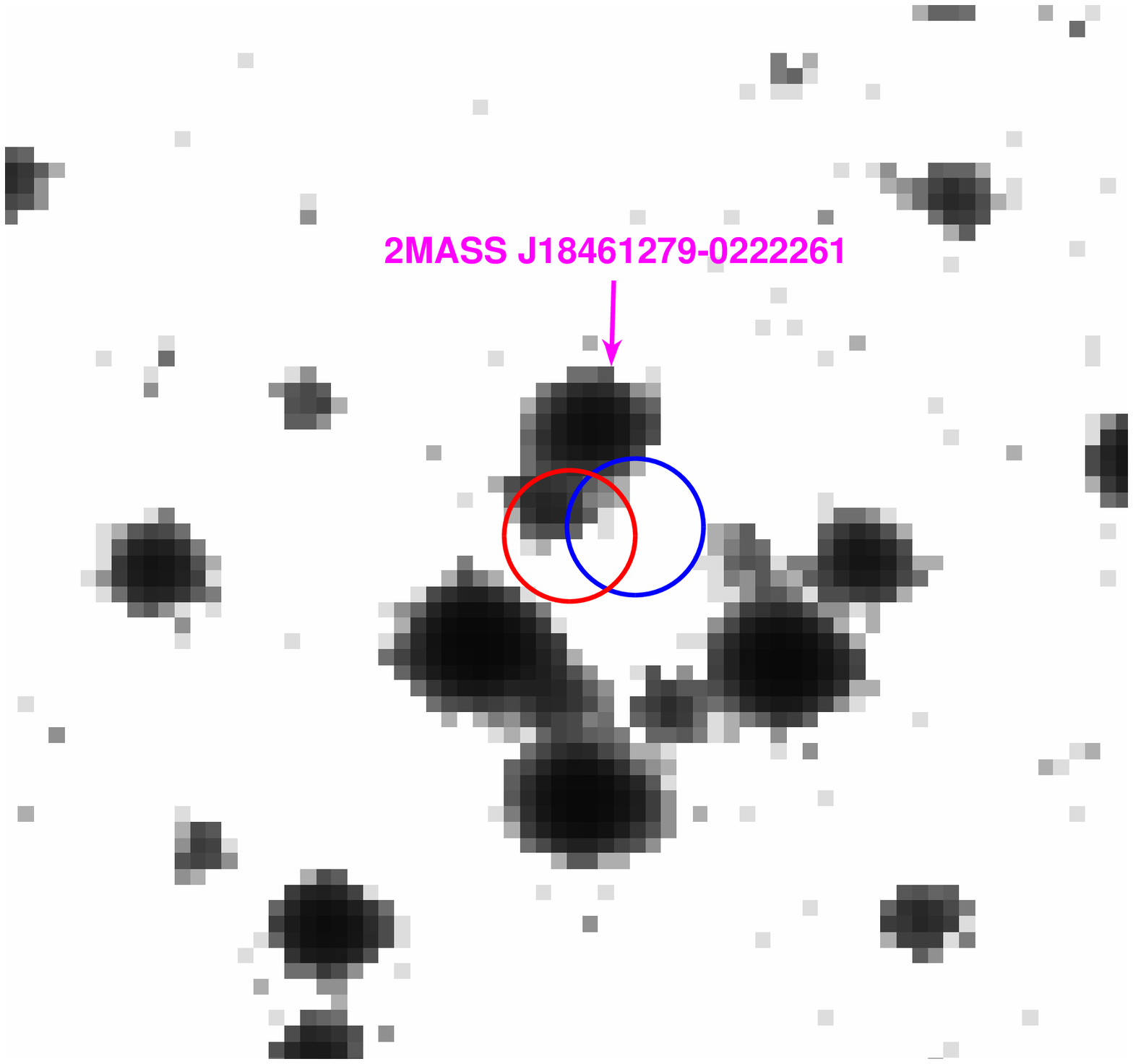}
\includegraphics[height=8.5cm,width=8.5cm,angle=0]{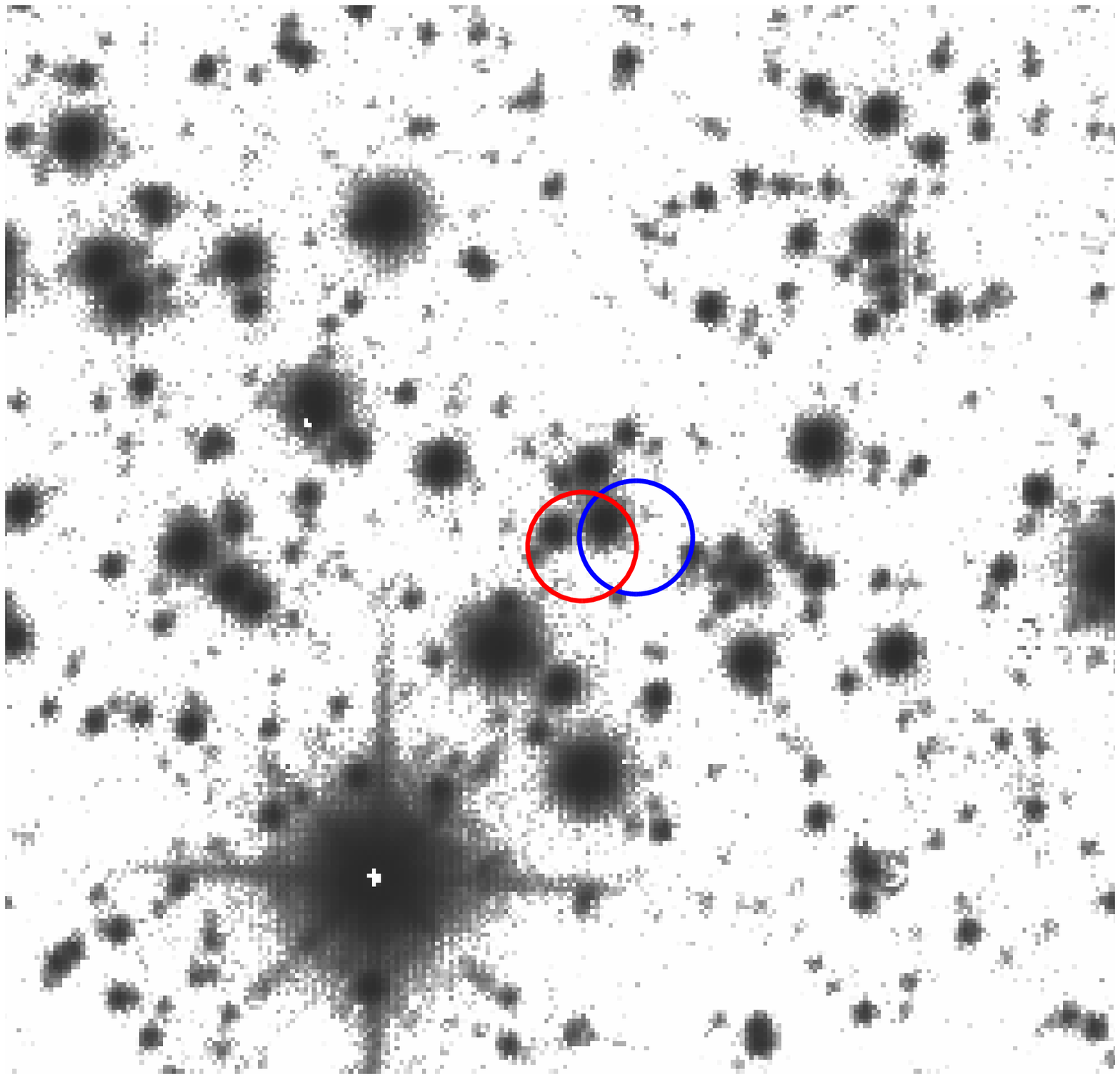}
\caption{$R$-band image acquired with the 'Cassini' telescope plus BFOSC 
(Upper panel) and UKIDSS $K$-band image (lower panel) of the field of 
IGR J18462$-$0223,  both with superimposed the 90\% 
confidence level {\it XMM-Newton} and {\it Swift} X--ray error circles 
(the right (blue) and the left (red) one, respectively).  For both images, 
the field size is about 40$''$$\times$40$''$; north is at top and east is to the left.
The 2MASS object mentioned by Bodaghee et al. (2012) is also labeled
and indicated with an arrow  in the upper panel, whereas the counterpart we 
propose in this paper is  the NIR object in the lower panel contained within 
the  intersection of the {\it Swift} and the {\it XMM-Newton} error circles.}
\end{center}
\end{figure} 

 \section{Summary and discussion}

We presented a comprehensive multiwavelength study of the poorly known candidate 
SFXT IGR J18462$-$0223, at X--rays in the energy bands 0.5--10 keV and 18--60 keV 
both in outburst and out-of-outburst state, as well as in optical and NIR bands.

The joint soft X--ray positional information obtained with {\it Swift/XRT} and {\it 
XMM-Newton}, together with optical and NIR observations allowed us to pinpoint the
very likely counterpart of this hard X--ray object as a very red source with
characteristics fully consistent with those of a heavily reddened early-type supergiant.
Assuming thus a blue supergiant nature for the secondary component of the X--ray binary
IGR J18462$-$0223, we can determine the reddening toward the source by considering
the intrinsic NIR colors of such stars (Wegner 1994). In the present case, we observe
$H-K$ = 2.54, whereas for early-type supergiants the intrinsic value of this color
is ($H-K)_0 \sim$ 0, which implies a color excess $E(H-K$) $\sim$ 2.5. Using the 
Milky Way extinction law of Cardelli et al. (1989) this means a reddening $A_V \approx$ 
40 mag, or $A_K \approx$ 4 mag. This is of course consistent with the nondetection of 
the object in the $R$ band, and also in the $J$ one given that a magnitude $J \sim$ 
22.5 is expected for the source in this scenario, thus well below the UKIDSS limit
in this band. 

This reddening, using the formula of Predehl \& Schmitt (1995), implies a column 
density $N_{\rm H} \sim$ 6.4$\times$10$^{22}$ cm$^{–2}$, which is larger than that of 
the Galaxy along the line of sight of the source, but smaller than that measured in
X--rays (see Table 2): this suggests that (i) the object is very far from Earth,
possibly on the far side of the Galaxy, and (ii) additional extinction is present 
in the vicinity of the accretor.
Concerning point (i) we can try to estimate the distance of IGR J18642$-$0223 again
in the assumption that it hosts an OB supergiant. Using $A_V \sim$ 40 mag and the 
tabulated absolute magnitudes (Lang 1992) and colours (Wegner 1994) for this type 
of star we find a distance of $\sim$11 kpc, thus consistent with a Scutum arm 
tangent location for the source as suggested by Bodaghee et al. (2012).

As regards the X--ray emission, and in particular  the X-ray outburst behavior, if we 
assume a  distance of $\sim$ 11 kpc  then  IGR J18462$-$0223 displays  peak-outburst X-ray luminosities in the range $\sim$ (4--22)$\times$10$^{36}$ erg  s$^{-1}$ (18--60 keV) or  
5.8$\times$10$^{36}$ erg  s$^{-1}$ (3--10 keV), i.e.  values similar   to those of  known confirmed SFXTs. Remarkably, 
the duration of all the X-ray  outbursts detected  by {\it INTEGRAL} is of the order of few hours, i.e. in 
the range 1--12 hours.  Grebenev \& Sunyaev (2010)  argued that this  characteristic   
could be due to an observational effect related to  the source  large distance  or alternatively  to its location at the edge of the {\it INTEGRAL}/IBIS
FoV when detected during the outbursts,  i.e. a region where the effective area of the telescope
decreases.  We tend to favor the former hypothesis since some outbursts have been detected by {\it INTEGRAL}/IBIS at low off-axis angle (e.g. $\sim$ 6$^\circ$). It is likely  that, due to the  relatively large distance of the source ($\sim$ 11 kpc),  only the brightest ($L_{{\rm X}}$  $\ge$ 10$^{36}$  erg s$^{-1}$) and shortest  (few hours duration) tops of the outbursts  are detectable while the  longer and lower  intensity X-ray outburst activity is just too faint to be detected  at energies above 18  keV. We  found that the time  IGR J18462$-$0223 spends in bright outbursts (as detected by {\it INTEGRAL}) is $\sim$ 1.2\% of the total. This value is  in the range  0.1\%--3\% found by Ducci et al. (2010) who analyzed {\it INTEGRAL} data  of a sample of 
firm/candidate SFXTs. 

From archival {\it INTEGRAL}/IBIS  observations,  we placed a 3$\sigma$ upper limit of $\sim$ 8$\times$10$^{-12}$ erg cm$^{-2}$  s$^{-1}$  
 (1.1$\times$10$^{35}$ erg s$^{-1}$ at 11 kpc) to the hard X-ray emission during the out-of-outburst state (18--60 keV),  this  is the most stringent constrain above 18 keV to date. Archival {\it ASCA} and {\it Chandra}  observations  also allowed us to infer  a deep 3$\sigma$ upper limit on the soft X-ray flux (0.5--10 keV) of the order of 2.9$\times$10$^{-13}$ erg cm$^{-2}$ s$^{-1}$ which translates into a luminosity of    $L_{{\rm X}}$$<$4.1$\times$10$^{33}$ erg s$^{-1}$ at 11 kpc. 
Such value implies a lower limit on the  dynamic range  equal to  $>$1,380 (0.5--10 keV). 
It is very likely that {\it ASCA} and {\it Chandra} observed the source during its  soft X-ray quiescence which is  a very rare state for SFXTs,  being characterized by no accretion,   luminosity values of  $L_{{\rm X}}$ $\sim$  10$^{32}$ erg  s$^{-1}$  and very soft X-ray spectrum 
(e.g. see int'Zand 2005).  Conversely, we reported a {\it Swift}/XRT  detection  with an absorbed  flux (luminosity) of 1.6 $\times$10$^{-11}$ erg cm$^{-2}$ s$^{-1}$ 
(2.3$\times$10$^{35}$ erg s$^{-1}$ at 11 kpc).  Such value is  at least two order of magnitudes higher than those  inferred from the {\it ASCA}  and  {\it Chandra} observations during the likely quiescence,  moreover  it is  about one--two order of magnitude lower than those typically measured during X-ray outbursts. It is likely that our reported  {\it Swift}/XRT detection of  the source, in addition to the {\it XMM} one reported by Bodaghee et al. 2012 with a similar flux,  represent  the so-called intermediate intensity  X-ray  state during which SFXTs spend the majority of their time (Sidoli et al. 2008) with typical $L_{{\rm X}}$ $\sim$ 10$^{34}$  erg  s$^{-1}$  and hard X-ray spectra ($\Gamma$$\sim$ 1--2).  During such intermediate X-ray state, SFXTs are still accreting material although at a much lower level  than that during the bright X-ray outbursts (Sidoli et al. 2008).   The hard X-ray spectrum  ($\Gamma$$\sim$1.5) and the X-ray flux values measured by both {\it Swift}/XRT and {\it XMM}  are compatible   with the intermediate intensity state scenario.

\begin{figure}
\begin{center}
\includegraphics[height=6cm,width=8cm]{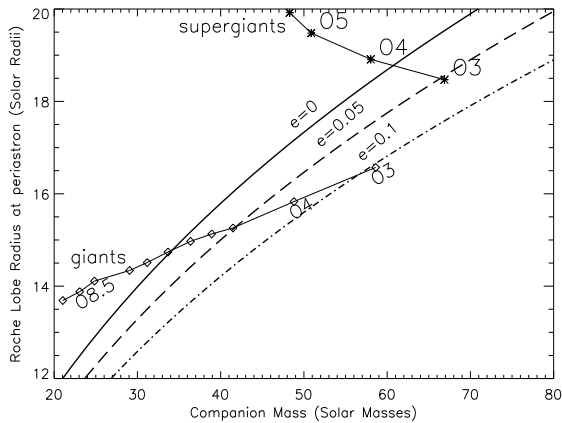}
\caption{Roche lobe radius at periastron versus the companion mass, assuming an orbital period of 2.13~days,
for different eccentricities (solid line: e=0; dashed line: e=0.05; dash-dotted line: e=0.1).
Radii and masses of O-type stars are also overplotted, taken from Martins et al. (2005):
diamonds mark O-type giant stars (luminosity class III),
while asterisks indicate O-type supergiants (luminosity class I), for different spectral sub-types.}
\end{center}
\end{figure} 

We found an hint of a possible  periodicity  of  $\sim$ 2.13   days since all five X-ray outbursts detected by {\it INTEGRAL}/IBIS to date  are spaced by 
multiples of 2.13 days. This periodicity  could be interpreted as the likely  orbital period of the binary system.  
Unfortunately  we cannot draw any firm conclusion on its  genuine existence  because no unambiguous and significant signal (e.g. significance $>$99.99\%) at  2.13 days has been found with a proper  periodicity analysis. In fact,  we note  that  
in  both {\it RXTE}/ASM periodograms obtained  with two different methods (Lomb--Scargle and fast Fourier transform) a 2.13 days  periodicity 
has been detected with a significance  too low ($\sim$ 90\%)  to  claim a secure and firm detection. 

Firstly, we note  that Bodaghee et al. (2012) have  recently detected X-ray pulsations at $\sim$ 997 seconds from IGR~J18462$-$0223. If we suppose that the tentative orbital period  of 2.13~days is real, then this would nicely place the system in the region of the Corbet diagram which host the wind--fed SGXBs and some other SFXTs (IGR~J17544$-$2619, Drave et al. 2012; IGR J16418$-$4532, Sidoli et al. 2012), 
 however the same holds also for significantly longer orbital periods, i.e. up to $\sim$ 30--40 days.

More importantly, if we suppose that the tentative orbital period  of 2.13 days is real, then  IGR~J18462$-$0223  would be the SFXT with the shortest  orbital period among the currently known systems.  Firstly, this scenario poses the question if 
a massive   early-type OB supergiant star is compatible or not with such a  short orbital period. 
Assuming that the compact object is a neutron star with a mass of 1.4~M$_{\odot}$,
we plot in Fig. 7 the Roche lobe radius of the companion (at periastron, if the orbit is not circular) versus its mass and  for different system eccentricities  (Eggleton 1983).  The radii of O type stars, I and III luminosity classes (Martins et al. 2005),  are overplotted for comparison.
From this plot it is evident that  a typical O type supergiant,  and hence  a SFXT nature, would be compatible with the 2.13~days orbital period.  
Also early type giant stars (earlier than O6, if the orbit is circular)  represent viable possibilities as 
donor stars in IGR~J18462$-$0223. 

In addition, it is worth noting that in the scenario of IGR~J18462$-$0223 having a short orbital period, 
its OB supergiant star would not necessarily fill its Roche lobe and other effects could also  take place. In fact as the supergiant  
donor  star expands and is close to fill  its Roche lobe, the wind could be not spherically symmetrical 
(as for isolated stars)  but strongly enhanced toward the compact object (focused wind), as e.g.  the case of the persistent SGXB Cyg X-1 
(Friend \& Castor 1982). Moreover Sidoli et al. (2012), using a 40 ks {\it XMM-Newton} observation,  recently proposed that the neutron 
star  hosted in the narrow orbit of the SFXT IGR~J6418$-$4532 ($\sim$ 3.7 days orbital period) could be accreting  in a regime that is transitional  between pure wind  and full Roche lobe overflow.

In such scenario the mass loss from the supergiant star is  
dominated by the strong wind, but with the important additional contribution of a tidal gas stream, focused towards the neutron star compact object. 
This mechanism produces extreme variations in the mass accretion rate (mainly due to the dynamical interaction of the weak tidal gas stream with the accretion bow shock around the neutron star) and is responsible for the observed  marked X-ray variability on short timescale.  Interestingly, 
Sidoli et al. (2012)   proposed  that such transitional Roche lobe overflow scenario  could  be a dominant process not only in the case of 
the SFXT IGR~J6418$-$4532  but also in other SFXTs  with similar short orbital periods,  e.g. our specific case of IGR~J18462$-$0223.  
To date, the only available long ($\sim$ 30 ks) soft X-ray observation ({\it XMM}) of  IGR~J18462$-$0223
was  recently reported by Bodaghee et al. (2012); the source was detected at an intermediate   flux level of 
$\sim$ 3$\times$10$^{-11}$ erg cm$^{-2}$ s$^{-1}$ (0.5--10 keV) with 
a  variability of an order of magnitude on timescales as short as $\sim$ 15 minutes. Unfortunately,  other
soft X-ray observations  are still lacking and are strongly needed to fully explore and consider whether or not a 
transitional Roche lobe overflow is occurring.  Presently  it is puzzling to explain how a SFXT with a short 
orbital period, like IGR~J18462$-$0223, could spend a considerable fraction of time at very low X-ray luminosities out-of-outbursts,  
as inferred especially  from INTEGRAL long term monitoring above 18 keV 
($L_{{\rm X}}$ $<$ 10$^{35}$ erg s$^{-1}$).  One possible explanation is  that some mechanisms could likely be at work 
to reduce/stop  the mass accretion rate onto the neutron star,  i.e. centrifugal and/or  magnetic barrier 
(Bozzo et al. 2008, Grebenev et al. 2007), transitions between two different regimes of plasma cooling (Shakura et al. 2012).

Further studies on  IGR~J18462$-$0223, especially a  long term monitoring across all orbital phases, 
would be very useful  to test  the above models.  Moreover, further investigations are strongly needed to fully 
confirm (or reject) the genuine existence of the putative 2.13 days periodicity. 
Also, NIR spectroscopy is needed to definitely confirm the nature of the reddened UKIDSS source presented
in this paper as the likely counterpart of IGR J18462$-$0223 and its association with this interesting hard X--ray object.

\begin{acknowledgements}
We are very grateful to the referee, Ignacio Negueruela, for several important comments which substantially
improved this paper. The italian authors acknowledge the ASI financial support via grant ASI-INAF I/033/10/0 and I/009/10/0.
This work was also supported by the grant from PRIN-INAF 2009 (The transient X-ray sky: new classes of X-ray binaries
containing neutron stars, PI: L. Sidoli).  S. P. Drave acknowledges support from the Science and Technology Facilities Council, STFC. This research has made use of the public access {\it RXTE}/ASM light curves provided by the ASM team (see 
http://xte.mit.edu/ASM$\textunderscore$lc.html). It has also made use of the VizieR catalogue access tool at CDS, Strasbourg, France. Based on observations collected at the Italian Telescopio 
Nazionale Galileo, located at the Observatorio del Roque de los Muchachos (Canary
Islands, Spain) and at the 'G.D. Cassini' telescope of the Astronomical
Observatory of Bologna in Loiano (Italy).  We  thank Vania Lorenzi for coordinating our Service Mode 
observations at TNG, and Silvia Galleti and Antonio De Blasi for the acquisition of images in Service Mode at Loiano.
\end{acknowledgements}

\end{document}